\documentclass[12pt, prd, showpacs,showkeys]{revtex4}
\usepackage{amssymb}
\usepackage{amsmath}
\usepackage{graphicx}

\input{tcilatex}

\begin{document}

\title{High-energy collisions inside black holes and their counterpart in
flat space-time}
\author{O. B. Zaslavskii}
\affiliation{Department of Physics and Technology, Kharkov V.N. Karazin National
University, 4 Svoboda Square, Kharkov 61022, Ukraine}
\email{zaslav@ukr.net }

\begin{abstract}
Two particles can collide inside a nonextremal black hole in such a way that
the energy $E_{c.m.}$ in their centre of mass frame becomes as large as one
likes. We show that this effect can be understood with the help of a simple
analogy with particle collisions in flat space-time. As the two-dimensional
part of near-horizon geometry inside a black hole is described by the flat
Milne metric, the results have a general character. Full classification of
scenarios with unbound $E_{c.m.}$ is suggested. Some scenarios of this kind
require proximity of collision to the bifurcation point, but for some other
ones this is not necessary.
\end{abstract}

\maketitle

\section{Introduction}

The Ba\~{n}ados-Silk-West effect (the BSW effect) consists in getting
arbitrarily large energies $E_{c.m.}$ in the centre of mass frame of
particles colliding near black holes \cite{ban}, i.e. in the strong
gravitational field. Nonetheless, it turned out that a kinematic nature of
the BSW effect outside the event horizon can be revealed in terms of a very
simple model of particles colliding in \textit{flat }space-time (see Sec. VI
of \cite{ah}). Now, we show the existence of analogy of this kind for
collisions inside black holes. (We do not touch upon other numerous aspects
of the BSW effect.) Thus for the BSW effect near the inner black hole
horizon \cite{gp-astro} - \cite{bif}, there exists its counterpart in the
flat space-time. One observer explains the BSW effect by the proximity to
the horizon and a special character of trajectories of colliding particles.
But another observer does not see the horizon at all, so his explanation
should be qualitatively different. Moreover, the two-dimensional part of
near-horizon geometry is described by the Milne metric (see below) which is
flat, so our consideration relies not on some particular model but has a
general character.

It was shown in \cite{inner}, \cite{bif} that one of colliding particles
should follow the trajectory that would pass through the bifurcation point
where the future and past horizons meet. As such a point does not show up in
realistic black holes, one could think that corresponding scenarios of
collision looked somewhat academic. Here, we show that, nonetheless, there
exists also such a scenario of the BSW effect in which the actual point of
collision may be near the horizon far from the bifurcation point. One can
hope that this makes the BSW effect near the inner black hole or
cosmological horizon more physical.

One reservation is in order. The whole space-time of an eternal black hole
consists of the "black hole" and "white hole" parts. I discuss motion of
particles from the inner non-static region to the outer static one that
corresponds to the white hole region. However, for brevity and in accordance
with tradition, I use the term "black hole" anyway. According to the
Novikov's classification \cite{nov}, this is the so-called T$_{+}$ region.
The pictures describing collision are presented on Fig. 1 and Fig. 2 where
the most interesting cases are drawn and the trajectories of the colliding
particles, horizons and bifurcation point O are indicated (see for details
below).

\section{Minkowski and Milne metrics}

Let us consider the metric of a spherically symmetric black hole%
\begin{equation}
ds^{2}=-fdt^{2}+\frac{dr^{2}}{f}+r^{2}(d\theta ^{2}+\sin ^{2}\theta d\phi
^{2})\text{.}
\end{equation}

Here, it is assumed that $g_{00}g_{11}=-1$ that does not affect the essence
of matter but simplifies formulas. It is implied that there is a horizon at $%
r=r_{h}$, so $f(r_{h})=0$. We are interested in the region inside the
horizon, so $r\leq r_{h}$. Near the horizon, we can exploit the Taylor
expansion%
\begin{equation}
f=f_{1}(r-r_{h})+O((r-r_{h})^{2})\text{,}
\end{equation}%
where $f_{1}>0$. Then, omitting the angular part of the metric irrelevant in
the present context, we have near the horizon%
\begin{equation}
ds^{2}\approx f_{1}(r_{h}-r)dt^{2}-\frac{dr^{2}}{f_{1}(r_{h}-r)}\text{.}
\end{equation}

In our region, the coordinate $r$ has a time-like character, $t$ is a
space-like. Making the substitution%
\begin{equation}
r_{h}-r=\frac{f_{1}}{4}\tilde{t}^{2}\text{, }t_{h}-t=\frac{2\tilde{x}}{f_{1}}%
\text{,}
\end{equation}%
we arrive at the metric%
\begin{equation}
ds^{2}=-d\tilde{t}^{2}+\tilde{t}^{2}d\tilde{x}^{2}\text{.}  \label{mln}
\end{equation}%
In other words, a nonextremal black hole metric inside the horizon can be
approximated in the near-horizon region by the so-called Milne metric (\ref%
{mln}). This is a counterpart to the well-known fact that outside the
horizon the metric of a nonextremal black hole can be approximated (in the
two-dimensional subspace) by the Rindler metric.

The metric (\ref{mln}) is flat. It can be obtained from the Minkowski one

\begin{equation}
ds^{2}=-dt^{2}+dx^{2}\text{,}  \label{min}
\end{equation}

with the help of the coordinate transformations%
\begin{equation}
x=\tilde{t}\sinh \tilde{x}\text{,}  \label{x}
\end{equation}%
\begin{equation}
t=\tilde{t}\cosh \tilde{x}\text{.}  \label{t}
\end{equation}

We are interested in the lower quadrant of the entire plane, where 
\begin{equation}
t<0,\left\vert x\right\vert <\left\vert t\right\vert .  \label{xt0}
\end{equation}%
The inverse transformation reads%
\begin{equation}
\tilde{t}^{2}=t^{2}-x^{2}\text{,}  \label{xyt}
\end{equation}%
\begin{equation}
\tanh \tilde{x}=\frac{x}{t}\text{.}  \label{ttx}
\end{equation}

There are two horizons in the metric (\ref{mln}): the right horizon $x=-t$,
where%
\begin{equation}
\tilde{t}=0\text{, }\tilde{x}=-\infty \text{,}  \label{r}
\end{equation}%
and the left one $x=+t$, where 
\begin{equation}
\tilde{t}=0\text{, }\tilde{x}=+\infty .  \label{l}
\end{equation}

There is also the "bifurcation point", where both horizons meet, so $x=0=t$,%
\begin{equation}
\tilde{t}=0\text{, }\left\vert \tilde{x}\right\vert <\infty .  \label{bif}
\end{equation}

This is point O on figs.1 and 2.

The metric (\ref{mln}) has a space-like Killing vector which can be written
in these coordinates as%
\begin{equation}
\xi ^{\mu }=(0,1)\text{.}
\end{equation}

The corresponding momentum $X=mu_{\mu }\xi ^{\mu }$ ($m$ is the particle's
mass) is conserved. Here, $u^{\mu }=\frac{dx^{\mu }}{d\tau }$ is the
four-velocity, $\tau $ is the proper time. It follows from the geodesic
equations (in which $m=1$) that%
\begin{equation}
\frac{d\tilde{x}}{d\tau }=\frac{X}{\tilde{t}^{2}}\text{,}  \label{1}
\end{equation}%
\begin{equation}
\frac{d\tilde{t}}{d\tau }=-\frac{Z}{\tilde{t}}>0\text{, }Z=\sqrt{X^{2}+%
\tilde{t}^{2}}\text{,}  \label{0}
\end{equation}%
so that%
\begin{equation}
\frac{d\tilde{x}}{d\tilde{t}}=-\frac{X}{Z\tilde{t}}\text{.}  \label{xtz}
\end{equation}

After integration, one finds that%
\begin{equation}
\tilde{t}=\frac{X}{\cosh \tilde{x}_{0}(V\cosh \tilde{x}-\sinh \tilde{x})}=%
\frac{x_{0}}{\sinh \tilde{x}-V\cosh \tilde{x}}\text{,}  \label{tx}
\end{equation}%
\begin{equation}
V=\tanh \tilde{x}_{0}\text{,}  \label{v0}
\end{equation}%
\begin{equation}
X=-x_{0}\cosh \tilde{x}_{0}\text{, }  \label{xv}
\end{equation}%
where $\tilde{x}_{0}$ and $x_{0\text{ }}$are constants. It follows from (\ref%
{xv}) that%
\begin{equation}
X=-x_{0}E\text{, }E=\frac{1}{\sqrt{1-V^{2}}}.  \label{xe}
\end{equation}

With the help of (\ref{x}), (\ref{t}), one can recognize in (\ref{tx}) the
standard equation of motion in the Minkowskian coordinates%
\begin{equation}
x-Vt=x_{0}\text{.}  \label{xvt}
\end{equation}%
Here, $V\,\ $has the meaning of velocity, $E$ being the energy.

The frame (\ref{mln}) coresponds to the observer (mentioned in\
Introduction) who sees the horizon while frame (\ref{min}) corresponds to
the observer who does not. In what follows, we discuss properties of
collisions in both Milne and Minlowski frames related by coordinate
transformations (\ref{x}), (\ref{t}) and (\ref{xyt}), (\ref{ttx}). It is
supposed that colliding particles follow geodesics which are described by
eq. (\ref{tx}) in the Milne frame and eq. (\ref{xvt}) in the Minkowski frame.

\section{Collision of two particles: general formulas}

Let particles 1 and 2 with the momenta $X_{1}$ and $X_{2}$, four-velocities $%
u_{1}^{\mu }$ and $u_{2}^{\mu }$, masses $m_{1}$ and $m_{2}$ collide. Then,
the energy in the centre of mass frame%
\begin{equation}
E_{c.m.}^{2}\equiv -(m_{1}u_{1}^{\mu }+m_{2}u_{2}^{\mu })(m_{1}u_{1\mu
}+m_{2}u_{2\mu })=m_{1}^{2}+m_{2}^{2}+2m_{1}m_{2}\gamma \text{,}  \label{ecm}
\end{equation}%
where the Lotentz factor of relative motion is

\begin{equation}
\gamma =-u_{1\mu }u_{2}^{\mu }\text{.}  \label{ga}
\end{equation}

Calculating $\gamma $ in the frame (\ref{mln}), one can obtain from (\ref{1}%
), (\ref{0}), (\ref{ga}) that 
\begin{equation}
\gamma =\frac{Z_{1}Z_{2}-X_{1}X_{2}}{\tilde{t}^{2}}\text{,}  \label{gat}
\end{equation}%
where $Z_{1,2}$ are given by eq. (\ref{0}), $m_{1}=m_{2}=1$ for simplicity.
In the Minkowski frame, where 
\begin{equation}
u^{\mu }=\frac{1}{\sqrt{1-V^{2}}}(1,V)\text{,}  \label{um}
\end{equation}

(\ref{ga}) can be rewritten as%
\begin{equation}
\gamma =E_{1}E_{2}(1-V_{1}V_{2})\text{.}  \label{gm}
\end{equation}

In the point of collision, 
\begin{equation}
\tilde{t}_{1}=\tilde{t}_{2}=\tilde{t}_{c}\text{, }\tilde{x}_{1}=\tilde{x}%
_{2}=\tilde{x}_{c},  \label{col}
\end{equation}%
we obtain from (\ref{tx}) that

\begin{equation}
\tanh \tilde{x}_{c}=\frac{[\left( x_{0}\right) _{2}V_{1}-\left( x_{0}\right)
_{1}V_{2}]}{[\left( x_{0}\right) _{2}-\left( x_{0}\right) _{1}]}.
\label{tanh}
\end{equation}

\section{Structure of the light cone and the BSW effect}

Calculation of the Lorentz factor $\gamma $ can be also described in the
geometric terms on the basis of the approach of \cite{bif} and \cite{cqg}.
Let us introduce in a given point two independent light-like basis vectors $%
l^{\mu }$ and $N^{\mu }$. It is convenient to normalize them according to $%
l^{\mu }N_{\mu }=-1$. The four-velocity $u_{i}^{\mu }$ of each particle ($%
i=1,2$) can be expanded as%
\begin{equation}
u_{i}^{\mu }=\beta _{i}N^{\mu }+\frac{l^{\mu }}{2\alpha _{i}}.
\end{equation}

Then, the quantity (\ref{ga}) can be written as%
\begin{equation}
\gamma =\frac{1}{2}(\frac{\beta _{1}}{\alpha _{2}}+\frac{\beta _{2}}{\alpha
_{1}})\text{.}  \label{ga1}
\end{equation}%
Here,%
\begin{equation}
\beta =-l_{\mu }u^{\mu }\text{, }\alpha =-\frac{1}{2}(N_{\mu }u^{\mu })^{-1}%
\text{.}  \label{ab}
\end{equation}

The choice of vectors $l^{\mu }$ and $N^{\mu }$ is ambiguous, one can change
them according to $l^{\mu }\rightarrow \lambda l^{\mu }$, $N^{\mu
}\rightarrow \lambda ^{-1}N^{\mu }$. However, this entails the
transformation $\beta _{i}\rightarrow \lambda \beta _{i}$, $\alpha
_{i}\rightarrow \lambda \alpha _{i}$, so that (\ref{ga1}) remains intact.
Let us choose the vectors according to 
\begin{equation}
\tilde{l}^{\mu }=(-\tilde{t}\text{, }1)\text{, }
\end{equation}%
\begin{equation}
\tilde{N}^{\mu }=\frac{1}{2}(-\frac{1}{\tilde{t}}\text{, }-\frac{1}{\tilde{t}%
^{2}})\text{.}
\end{equation}%
It is convenient to introduce light-like coordinates 
\begin{equation}
\eta =t+x,\zeta =t-x.  \label{null}
\end{equation}%
Then, the same vectors in the Minkowski frame are equal to%
\begin{equation}
l^{\mu }=\zeta (-1,1),
\end{equation}%
\begin{equation}
N^{\mu }=\frac{1}{2\zeta }(-1,-1),
\end{equation}%
where we used (\ref{xyt}) in the calculation of $N^{\mu }$.

Calculating the coefficients with the help of eqs. (\ref{1}), (\ref{0}), we
obtain that $\beta =\alpha $, where 
\begin{equation}
\beta =Z-X\text{.}  \label{ba}
\end{equation}

In the Minkowski frame, using (\ref{um}) one finds a simple expression for $%
\beta $ in terms of velocity:%
\begin{equation}
\beta =-\frac{\zeta (1+V)}{\sqrt{1-V^{2}}}=\zeta \sqrt{\frac{1-\left\vert
V\right\vert }{1+\left\vert V\right\vert }}\text{.}  \label{b}
\end{equation}

\section{Near-horizon collisions}

With these general formulas at hand, we are already able to analyze the
conditions that lead to the BSW effect. In what follows, we call a particle
critical if $X=0$ and usual if $X\neq 0$. If $X$ does not vanish precisely
but is small, we call a particle near-critical. We are interested in the
cases when $\gamma $ in (\ref{ecm}) can become unbound despite finite $%
X_{1,2}$. It is seen from (\ref{gat}) that the only possibility of getting
unbound $\gamma $ is to arrange collision with small $\tilde{t}$. Then,
either collision occurs near the bifurcation point (\ref{bif}) or near the
horizon (say, the right one) (\ref{r}). We discuss different cases depending
on $X_{i}$ and will show which of them correspond to the bifurcation point
or the horizon.

\subsection{$X_{1}X_{2}<0$}

For small $\tilde{t}_{c}$, it follows from (\ref{0}), (\ref{gat}) that%
\begin{equation}
\gamma \approx \frac{2\left\vert X_{1}X_{2}\right\vert }{\tilde{t}_{c}^{2}}%
\text{.}  \label{12t}
\end{equation}

As we try to arrange the conditions for the BSW effect, we assume that $X_{1}
$ and $X_{2}$ are separated from zero, so both particles are usual. It would
seem that the BSW effect occurs according to (\ref{12t}). However, more
careful inspection shows that this is not necessary so (see below). Small $%
\tilde{t}_{c}$ imply either the vicinity of the right or left horizon (then, 
$\left\vert \tilde{x}_{c}\right\vert \rightarrow \infty $ according to (\ref%
{r}), (\ref{l})) or the vicinity of the bifurcation point (\ref{bif}) ($%
\tilde{x}_{c}$ is finite). We will consider these cases separately.

\subsubsection{Collision near the horizon}

Let us try to arrange collision near, say, the right horizon. It follows
from (\ref{r}) and (\ref{tanh}) that 
\begin{equation}
V_{1}=-1+(1+V_{2})\frac{\left( x_{0}\right) _{1}}{\left( x_{0}\right) _{2}}%
\text{.}  \label{v12}
\end{equation}%
As, by assumption, $X_{1}$ and $X_{2}$ have different signs, $\left(
x_{0}\right) _{1}$ and $\left( x_{0}\right) _{2}$ also have different signs
according to (\ref{xe}). Then, it follows from eq. (\ref{v12}) that $%
V_{1}<-1 $ that is impossible. (In a similar way, near the left horizon we
would obtain $V_{1}>1$.) Thus the condition of collision (\ref{tanh}) is
inconsistent with $\left\vert V_{1}\right\vert <1$. Therefore, collision on
the horizon cannot occur. In the intermediate point with $\tilde{t}_{c}\neq
0 $ it is possible, but $\gamma $ is finite there, so there is no the BSW
effect. This is in agreement with previous studies \cite{gp-astro} - \cite%
{bif}.

\subsubsection{Scenario A: collision near the bifurcation point}

Near the bifurcation point, $t$ and $x$ are small, so according to (\ref{xvt}%
), $\left( x_{0}\right) _{1}$ and $\left( x_{0}\right) _{2}$ are also small.
As by assumption, $X_{1}$ and $X_{2}$ are separated from zero, (\ref{xe})
entails that $E_{1,2}\rightarrow \infty $, so $\left\vert V_{1,2}\right\vert
\rightarrow 1$. As is explained above, $\left( x_{0}\right) _{1}$ and $%
\left( x_{0}\right) _{2}$ have different signs. Then, by substitution into (%
\ref{tanh}), we see that if $V_{1}\approx 1\approx V_{2}$ or $V_{1}\approx
-1\approx V_{2}$, it turns out that $\left\vert \tilde{x}_{c}\right\vert
\rightarrow \infty $ contrary to the property (\ref{bif}). However, it is
possible to have $V_{1}\approx -V_{2}\approx \pm 1$. As in the quadrant
under discussion inequality (\ref{xt0}) holds, it is clear from (\ref{xvt})
that for $V_{i}$ $\approx +1$, we have $\left( x_{0}\right) _{i}>0$, \ so $%
X_{i}<0$ from (\ref{xe}) ($i=1,2$). In a similar way, for $V_{i}$ $\approx
-1 $, we have $\left( x_{0}\right) _{i}<0$ and $X_{i}>0$. Considering both
subcases, we obtain in the point of collision:

a) $V_{1}\approx +1$, $V_{2}\approx -1$, $X_{1}<0$, $X_{2}>0$.

Then, in the limit $\tilde{t}_{c}\rightarrow 0$, one obtains from (\ref{b})
that%
\begin{equation}
\beta _{1}\approx 2\left\vert X_{1}\right\vert \text{, }\beta _{2}\approx 
\frac{\tilde{t}_{c}^{2}}{2X_{2}}\text{.}  \label{b12}
\end{equation}%
From (\ref{ga1}) and (\ref{b12}), eq. (\ref{12t}) is recovered. From the
other hand, (\ref{gm}) gives us 
\begin{equation}
\gamma \approx 2E_{1}E_{2}\text{.}  \label{g12}
\end{equation}

Now, $\left( x_{0}\right) _{1}\left( x_{0}\right) _{2}\approx x^{2}-t^{2}=-%
\tilde{t}^{2}$, where we used (\ref{xyt}). Therefore, (\ref{12t}) agrees
with (\ref{g12}). Here, both $E_{1}$ and $E_{2}$ grow unbound. However, $%
X_{1}$ and $X_{2}$ are finite!

b) $V_{1}\approx -1$, $V_{2}\approx +1,X_{1}>0$, $X_{2}<0$,%
\begin{equation}
\beta _{2}\approx 2\left\vert X_{2}\right\vert \text{, }\beta _{1}\approx 
\frac{\tilde{t}_{c}^{2}}{2X_{1}}\text{,}
\end{equation}%
eq. (\ref{g12}) holds.

\subsection{$X_{1}X_{2}\geq 0$}

First, let both $X_{1}$ and $X_{2}$ be separated from zero (both particles
are usual). We are interesting in the case of small $\tilde{t}$ only. Then,
it is seen from (\ref{0}) that $Z_{1,2}\approx \left\vert X_{1,2}\right\vert
+\frac{\tilde{t}^{2}}{2\left\vert X_{1,2}\right\vert }$. As a consequence,
the gamma factor in (\ref{gat}) is finite, so there is no the BSW effect.

For $\gamma $ to be unbound, it is necessary that (i) collision occur for
small $\tilde{t}_{c}$, (ii) one particle (say, particle 1) be critical or
near-critical and particle 2 be finite. For definiteness, we take particle 1
to be precisely critical, so 
\begin{equation}
X_{1}=0.  \label{x10}
\end{equation}%
It is seen from (\ref{xv}) that in this case 
\begin{equation}
\left( x_{0}\right) _{1}=0.  \label{01}
\end{equation}%
Then, one can obtain from (\ref{xyt}), (\ref{xvt}), (\ref{tanh}) 
\begin{equation}
x_{c}=V_{1}t_{c}  \label{xv0}
\end{equation}%
\begin{equation}
V_{1}=\tanh \tilde{x}_{c}\text{,}  \label{v1c}
\end{equation}%
\begin{equation}
\tilde{t}_{c}=t_{c}\sqrt{1-V_{1}^{2}}\text{,}  \label{tt}
\end{equation}

\begin{equation}
X_{2}=\left\vert t_{c}\right\vert (V_{1}-V_{2})E_{2}=\frac{\left\vert \tilde{%
t}\right\vert (V_{1}-V_{2})E_{2}}{\sqrt{1-V_{1}^{2}}}\text{.}  \label{x20}
\end{equation}%
Here, according to (\ref{null}), (\ref{xvt}) and (\ref{01}), $\zeta
_{c}=t_{c}(1-V_{1})$. Then, it follows from (\ref{ba}), (\ref{b}) and (\ref%
{x20}) that%
\begin{equation}
\beta _{1}=\left\vert \tilde{t}_{c}\right\vert =\frac{X_{2}\sqrt{1-V_{1}^{2}}%
}{(V_{1}-V_{2})E_{2}}.  \label{b1t}
\end{equation}%
\begin{equation}
\beta _{2}=\left\vert \tilde{t}_{c}\right\vert \sqrt{\frac{1-V_{1}}{1+V_{1}}}%
(1+V_{2})E_{2}=\frac{X_{2}(1-V_{1})(1+V_{2})}{V_{1}-V_{2}}  \label{b2v}
\end{equation}%
Eq. (\ref{gat}) simplifies to%
\begin{equation}
\gamma =\frac{Z_{2}}{\left\vert \tilde{t}_{c}\right\vert }\text{.}
\label{gz}
\end{equation}

Below, we list the results of analysis and enumerate different cases that
give rise to unbound $\gamma $. We indicate the relevant quantities in terms
of both frames (\ref{min}) and (\ref{mln}). In all cases, $\tilde{t}%
\rightarrow -0$, $X_{2}\neq 0$, $\beta _{1}$ is given by eq. (\ref{b1t}),
eq. (\ref{gz}) gives us%
\begin{equation}
\gamma \approx \frac{X_{2}}{\left\vert \tilde{t}_{c}\right\vert }\text{.}
\label{gc}
\end{equation}

\subsubsection{Scenario B: collision near the bifurcation point}

According to (\ref{bif}), near the bifurcation point $\tilde{x}_{c}$ is
finite. Therefore, it follows from (\ref{v1c}) that $\left\vert
V_{1}\right\vert <1$. For a usual particle, eq. (\ref{x20}) holds. It is
seen from it that, to reconcile finite nonzero $X_{2}$ with small $x_{c}$,
we must have large $E_{2}\ $\ Then, it follows from (\ref{xe}) \ that $%
\left\vert V_{2}\right\vert \rightarrow 1$. There are two subcases depending
on the sign of $V_{2}$.

a) $V_{2}\approx -1.$ Then, it follows from (\ref{x20}) that $X_{2}>0$. It
is seen from (\ref{0}), (\ref{ba}) and (\ref{b2v}) that%
\begin{equation}
\beta _{2}\approx \frac{\tilde{t}_{c}^{2}}{2X_{2}}\approx \frac{%
X_{2}(1-V_{1})(1+V_{2})}{1+V_{1}}\text{,}
\end{equation}%
so in (\ref{ga1}), (\ref{gm}) 
\begin{equation}
\gamma \approx \frac{\sqrt{1+V_{1}}}{\sqrt{2}\sqrt{1-V_{1}}\sqrt{1+V_{2}}}%
\rightarrow \infty \text{.}
\end{equation}

b) $V_{2}\approx +1$. Then, it follows from (\ref{x20}) that $X_{2}<0$. Eqs.
(\ref{ba}) - (\ref{b2v}) give us 
\begin{equation}
\beta _{1}\approx \frac{\left\vert X_{2}\right\vert }{E_{2}}\sqrt{\frac{%
1+V_{1}}{1-V_{1}}}
\end{equation}%
\begin{equation}
\beta _{2}\approx 2\left\vert X_{2}\right\vert \text{.}  \label{b2}
\end{equation}%
whence%
\begin{equation}
\gamma \approx \frac{\sqrt{1-V_{1}}}{\sqrt{2}\sqrt{1+V_{1}}\sqrt{1-V_{2}}}%
\rightarrow \infty \text{.}
\end{equation}

\subsubsection{Scenario C: collision near the horizon}

Near the horizon, $\left\vert \tilde{x}_{c}\right\vert =\infty $ according
to (\ref{r}), (\ref{l}). Then, it follows from (\ref{v1c}) that $\left\vert
V_{1}\right\vert \rightarrow 1$. The value of $t_{c}$ is, generally
speaking, separated from zero, $\tilde{t}_{c}$ in (\ref{tt}) is small due to
the second factor. Therefore, collision can occur far from the bifurcation
point. It follows from (\ref{x20}) with finite $X_{2}$ that $E_{2}$ is
finite, so $\left\vert V_{2}\right\vert <1$. Now, there are two subcases
depending on the sign of $V_{1}$.

a) $V_{1}\approx -1$, $\tilde{x}_{c}\rightarrow -\infty $. Then, it follows
from (\ref{x20}) that $X_{2}<0$. According to (\ref{xv0}), collision occurs
near the right horizon $x=-t$. We see from (\ref{b1t}) that%
\begin{equation}
\beta _{1}\approx \frac{\sqrt{2}\left\vert X_{2}\right\vert \sqrt{1+V_{1}}%
\sqrt{1-V_{2}}}{\sqrt{1+V_{2}}}\text{,}
\end{equation}%
for $\beta _{2}$ eq. (\ref{b2}) holds.

Then, (\ref{ga1}) gives us%
\begin{equation}
\gamma \approx \frac{\sqrt{1+V_{2}}}{\sqrt{2}\sqrt{1+V_{1}}\sqrt{1-V_{2}}}%
\rightarrow \infty \text{.}
\end{equation}

b) $V_{1}\approx +1$, $\,$ $\tilde{x}_{c}\rightarrow +\infty $, now $X_{2}>0$
and 
\begin{equation}
\beta _{1}\approx \frac{\sqrt{2}X_{2}\sqrt{1-V_{1}}\sqrt{1+V_{2}}}{\sqrt{%
1-V_{2}}}
\end{equation}%
\begin{equation}
\beta _{2}\approx \frac{\tilde{t}_{c}^{2}}{2X_{2}}\approx \frac{%
(1+V_{2})(1-V_{1})X_{2}}{(1-V_{2})}\text{.}
\end{equation}%
\begin{equation}
\gamma \approx \frac{\sqrt{1-V_{1}}}{\sqrt{2}\sqrt{1+V_{2}}\sqrt{1-V_{1}}}%
\rightarrow \infty \text{.}
\end{equation}

For completeness, we discuss briefly the remaining cases. If $X_{1}=0=X_{2}$
(two critical particles), it is seen from (\ref{gat}) that $\gamma =1$, so
this degenerate case is uninteresting. Also, there is no BSW effect if $%
X_{i}=O(\tilde{t}_{c})$ (both particles are near-critical).

\subsection{Collision of near-critical and usual particles}

In scenarios B and C, we considered the case when one particle is precisely
critical. If particle 1 is near-critical and particle 2 is usual, it follows
from (\ref{gat}) that in the limit $\tilde{t}_{c}\rightarrow 0$,%
\begin{equation}
\gamma =\frac{1}{2}(\left\vert \frac{X_{1}}{X_{2}}\right\vert +\left\vert 
\frac{X_{2}}{X_{1}}\right\vert )\text{.}
\end{equation}

Taking also into account the above results for scenarios with collisions
near the horizon, one can write%
\begin{equation}
\lim_{X_{1}\rightarrow 0}\lim_{\tilde{t}_{c}\rightarrow 0}\gamma =\lim_{%
\tilde{t}_{c}\rightarrow 0}\lim_{X_{1}\rightarrow 0}\gamma =\infty \text{.}
\end{equation}

In other words, in scenarios B and C the BSW effect can be realized in two
basic ways: either (i) particle 1 is critical, collision happens near the
horizon (see Fig. 1) or (ii) collision happens on the horizon itself,
particle 1 is near-critical (see Fig. 2).

For collision on the horizon (say, the right one), it is seen from (\ref{xvt}%
) with $x=t$ that $\left( x_{0}\right) _{i}=t(1-V_{i})$. As $V_{i}<1$ and $%
t<0$, we have $\left( x_{0}\right) _{i}<0$ and, according to (\ref{xv}), $%
X_{i}>0$. In a similar way, $X_{i}<0$ for both particles if collisions occur
on the left horizon. Thus $X_{1}X_{2}>0$ for collision near the horizon.
There is no such a restriction for collisions near the bifurcation point.

\section{Summary of relevant scenarios}

In Table 1 below, we summarize the results of analysis.

Table 1. Types of collisions leading to the BSW effect for the Milne model.

\begin{tabular}{|l|l|l|l|l|l|l|l|}
\hline
Scenario & $V_{1}$ & $V_{2}$ & $X_{1}$ & $X_{2}$ & $\beta _{1}$ & $\beta
_{2} $ & Location \\ \hline
Aa & $\approx +1$ & $\approx -1$ & $<0$ & $>0$ & $2\left\vert
X_{1}\right\vert $ & $\sim \tilde{t}_{c}^{2}\ll \beta _{1}$ & Bifurcation
point \\ \hline
Ab & $\approx -1$ & $\approx +1$ & $>0$ & $<0$ & $\sim \tilde{t}_{c}^{2}$ & $%
2\left\vert X_{2}\right\vert \gg \beta _{1}$ & Bifurcation point \\ \hline
Ba & intermediate & $\approx -1$ & $0$ & $>0$ & $\sim \left\vert \tilde{t}%
_{c}\right\vert $ & $\sim \tilde{t}_{c}^{2}\ll \beta _{1}$ & Bifurcation
point \\ \hline
Bb & intermediate & $\approx +1$ & $0$ & $<0$ & $\sim \left\vert \tilde{t}%
_{c}\right\vert $ & $2\left\vert X_{2}\right\vert \gg \beta _{1}$ & 
Bifurcation point \\ \hline
Ca & $\approx -1$ & intermediate & $0$ & $<0$ & $\sim \left\vert \tilde{t}%
_{c}\right\vert $ & $2\left\vert X_{2}\right\vert \gg \beta _{1}$ & Horizon
\\ \hline
Cb & $\approx +1$ & intermediate & $0$ & $>0$ & $\sim \left\vert \tilde{t}%
_{c}\right\vert $ & $\sim \tilde{t}_{c}^{2}\ll \beta _{1}$ & Horizon \\ 
\hline
\end{tabular}

We indicate only scenarios that obey two criteria: (i) $\gamma $ is unbound,
(ii) two particle indeed meet in the same point, so eq. (\ref{col}) is
satisfied. To simplify presentation, we assumed that particle 1 in scenarios
B and C is exactly critical. This table gives us the full list of cases when
the Lorentz factor of relative motion $\gamma $ grows unbound for finite $%
X_{i}$. Strictly speaking, there is one more case not shown in the table
since it is trivial. It corresponds to infinitely large momenta $X_{i}$. In
scenario A, both particles are usual. In all other scenarios particle 1 is
critical (or near-critical), particle 2 is usual.

In all cases, according to (\ref{ga1}), $\gamma \approx \frac{1}{2}\frac{%
\beta _{2}}{\beta _{1}}$, if $\beta _{2}\gg \beta _{1}$ and $\gamma \approx 
\frac{1}{2}\frac{\beta _{1}}{\beta _{2}}$, if $\beta _{1}\gg \beta _{2}$.

The fact that in scenario Bb the coefficient $\beta _{1}\rightarrow 0$ for
the critical particle and $\beta _{2}\neq 0$ for a usual one is in agreement
with general properties of the bifurcation point relevant in the context of
the BSW effect \cite{bif}. Meanwhile, table 1 contains also some new
possibilities. In scenario Ba, both $\beta _{1}$ and $\beta _{2}$ vanish as
the bifurcation point is approached. However, this occurs with essential
different rates, so $\gamma $ becomes unbound according to (\ref{ga1}). In
scenario C, $t_{c}$ and $x_{c}\approx -t_{c}$ do not vanish, so collision
happens far from the bifurcation point.

If, instead of collisions in flat space-time (which turned out to be useful
methodical tool), we consider true black holes posesing the electric charge
or angular momentum, $X_{i}$ are not conserved \cite{inner}. Classification
itself retains its validity, provided $X_{i}$ are taken in the vicinity of
the horizon. In a general form, omitting details, the corresponding table
has the following structure.

Table 2. General types of collisions leading to the BSW effect inside black
holes.

\begin{tabular}{|l|l|l|l|l|}
\hline
Scenario & particle 1 & particle 2 & $X_{1}X_{2}$ & Location \\ \hline
A & usual & usual & $<0$ & Bifurcation point \\ \hline
B & critical or near-critical & usual & $0$ or small & Bifurcation point \\ 
\hline
C & critical or near-critical & usual & $0$ or $>0$ small & Horizon \\ \hline
\end{tabular}

\subsection{Comparison with previous studies}

It is instructive to compare the present results with those in our previous
studies \cite{inner}, \cite{bif}. Scenario A was considered in \cite{inner}
where it was displayed on Fig. 7. Scenario Bb was discussed in \cite{bif},
so case Ba extends the list of possibilities. Scenario C was actually
mentioned earlier in Sec. V B 2 of \cite{inner} (displayed on Fig. 7 there)
but identified not quite accurately in that, actually, one particle should
be near-critical, not simply usual.

Now, having sorted out all possible cases systematically, we obtained the
comprehensive list of relevant scenarios.

\section{Discussion and conclusion}

We have managed to construct classification of all possible scenarios of
collisions of particles near the inner horizon leading to the indefinite
growth of $E_{c.m.}$. This is done with the help of an elementary model of
two particles collision in flat space-time. In the Milne frame, the BSW
effect happens due to the existence of the horizon and a special character
of trajectories. In the Minkowski frame, there is no horizon. There, the
effect is due to the fact that a rapid particle hits a slow or motionless
one (or there is head-on collision of two rapid particles like in Scenario
A).

In true black hole metrics the whole situation is more complicated. However,
locally, in the small vicinity of the horizon from inside, the Milne metric
can be considered as a good approximation to the black hole metric.
Therefore, our consideration is quite generic and is applicable to the BSW
effect inside true black holes. In this context, "trivializing" the effect
by appealing to the Minkowski space-time, tangent to a given point, is a
tool to understand the BSW effect inside black holes where both the effect
itself and its properties were not so obvious (see discussion in \cite%
{gp-astro} - \cite{bif}). In other words, it is the simplified Milne metric
that enabled us to construct Table 2 of possible scenarios in a general case.

Scenario C of collisions tells us that the high energy collision can occur
far from the bifurcation point, its very existence is not mandatory at all.
This makes this version of the BSW effect more physical: it can manifest
itself any time a particle with the fine-tuned parameters crosses the
horizon in the direction from the inner nonstatic region to the outer static
one, be it the inner black hole horizon, the cosmological or the isolated
one \cite{is}.

\newpage Figures 
\begin{figure}[!ht]
\begin{center}
\includegraphics[width=8.6cm]{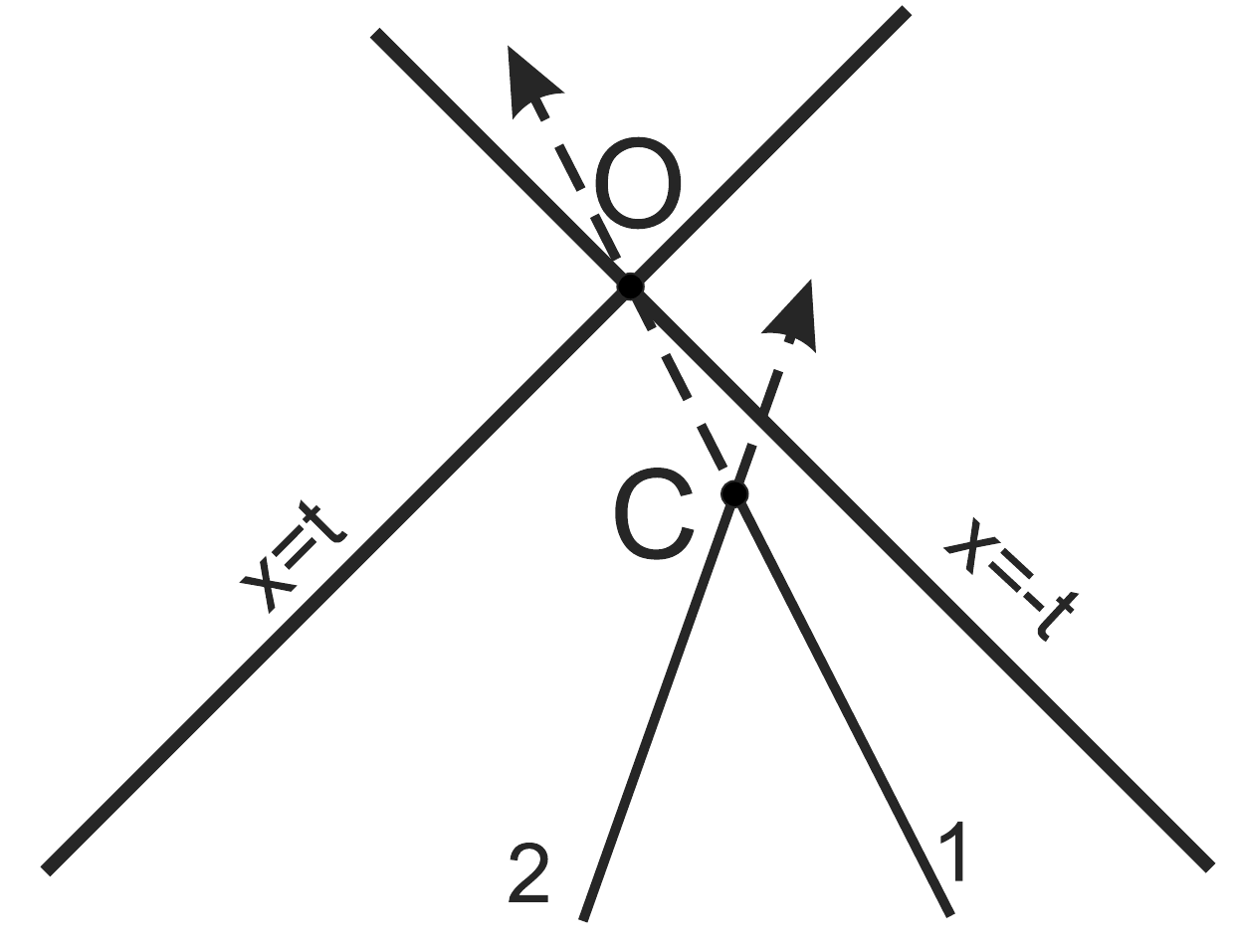}
\end{center}
\caption{Collision between the critical particle 1 and a usual particle 2
near the horizon.}
\label{Fig1}
\end{figure}

\begin{figure}[!hb]
\includegraphics[width=8.6cm]{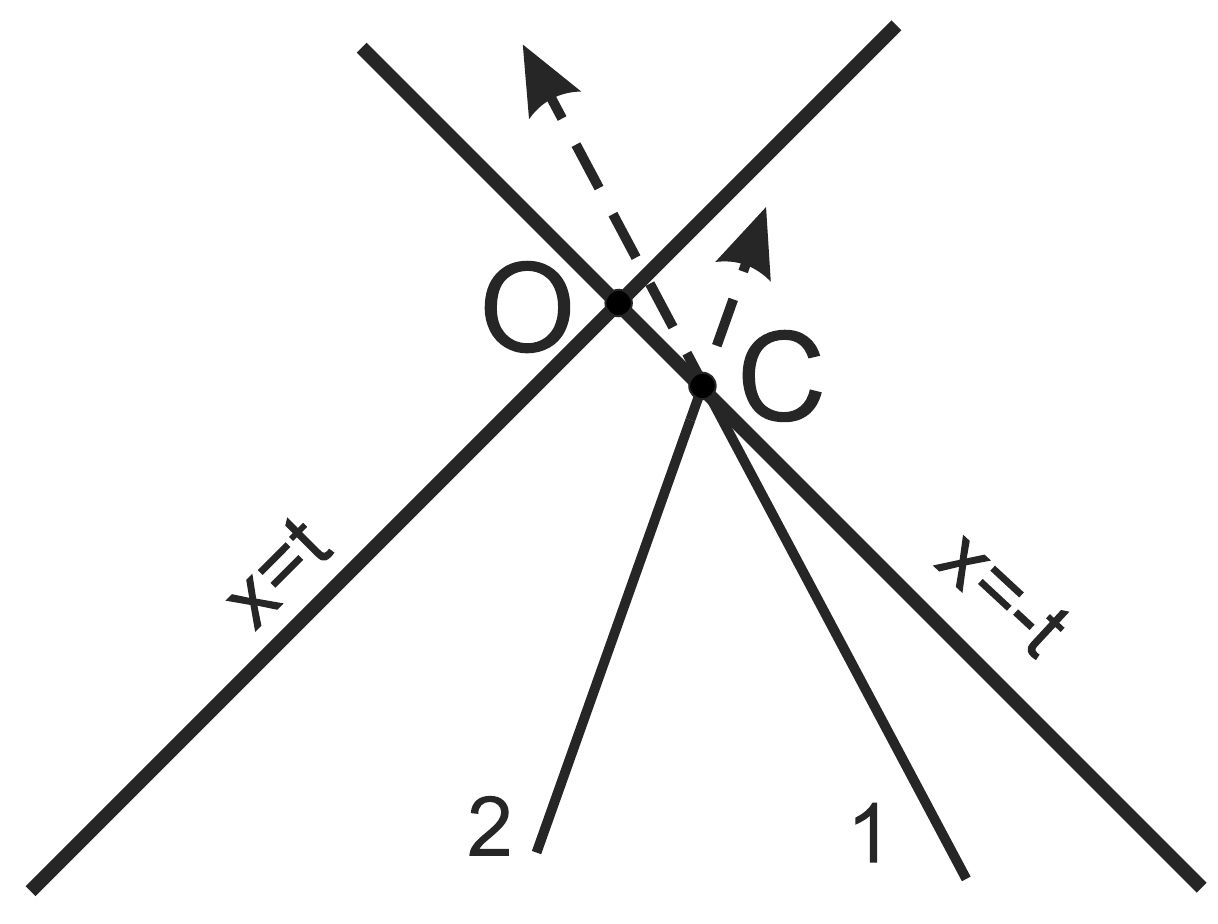}
\caption{Collision between the near-critical particle 1 and a usual particle
2 on the horizon.}
\label{Fig2}
\end{figure}

\end{document}